\documentclass{ws-procs961x669}  
\usepackage{latexsym,comment,verbatim}
\usepackage{amssymb}
\usepackage{amsfonts}
\usepackage{amsmath,color}

\def\mb#1{\mathbf{#1}}

\def\tt#1{\texttt{#1}}

\def\ber{\begin{eqnarray}}
\def\eer{\end{eqnarray}}
\def\beq{\begin{equation}}
\def\eeq{\end{equation}}

\def\ed{\end{document}}

\def\dT#1{\frac{\mathrm{d} #1}{\mathrm{d}T}}
\def\dTT#1{\frac{\mathrm{d} ^{2}#1}{\mathrm{d}T^{2}}}
\def\sT{\sin \left(\omega T \right)}
\def\cT{\cos \left(\omega T \right)}

\begin{document}
\title{Gravitomagnetic resonance and gravitational waves}

\author{Matteo Luca Ruggiero$^*$}

\address{Politecnico di Torino, Torino - Italy and INFN, Laboratori Nazionali di Legnaro, Legnaro - Italy\\
$^*$matteo.ruggiero@polito.it}

\author{Antonello Ortolan}
\address{INFN, Laboratori Nazionali di Legnaro, Legnaro - Italy}

\begin{abstract}
We show that using Fermi coordinates it is possible to describe the gravitational field of a wave using a gravitoelectromagnetic analogy. In particular, we show that using this approach, a new phenomenon,  called gravitomagnetic resonance, may appear. We describe it both from  classical and quantum viewpoints, and suggest that it could in principle be used as the basis for a new type of gravitational wave detectors.
\end{abstract}


\bodymatter

\section{Introduction}\label{aba:sec1}

General Relativity (GR) is the best model that  we have to understand gravitational interactions, and its predictions were verified with great accuracy during last century, even though we know that there are difficulties to explain, in the general relativistic framework, observations on galactic and cosmological scales, without claiming the existence of \textit{dark components}.\cite{2014arXiv1409.7871W,Debono:2016vkp} Remarkably, GR not only predicts corrections to known Newtonian effects such as in the case of perihelion advance, but there are general relativistic effects without Newtonian counterparts: for instance, this is the case of gravitational waves  and the so-called {gravitomagnetic  effects}  produced by mass currents.

As for gravitational waves, the first indirect evidence of their existence came from the observation of the binary pulsar B1913+16, whose orbital parameters are modified by the emission of gravitational waves.\cite{Hulse:1974eb,Weisberg:2004hi} It took about 100 years after the publication of Einstein's theory of gravity to obtain, in 2015, the first direct detection of gravitational waves,\cite{abbott2016observation} which was the beginning of gravitational wave astronomy. 

It is well known\cite{Ruggiero:2002hz} that  Einstein equations, in weak-field approximation (small masses, low velocities), can be written in analogy with Maxwell equations for the electromagnetic field, where the mass density and current play the role of the charge density and current, respectively; more in general, both the inertial and curvature effects in the vicinity of a given world-line, can be dealt with using a gravitoelectromagnetic formalism.\cite{Mashhoon:2003ax,Ruggiero_2020} These gravitomagnetic effects are very small if compared to the gravitoelectric ones, originating from mass density and, consequently, it is very difficult to measure them. Nonetheless, there have been various attempts and proposals: we remember  the  LAGEOS tests around the Earth,\cite{ciufolini2004confirmation,ciufolini2010gravitomagnetism} the subsequent LARES mission, \cite{ciufolini2012testing,ciufolini2016test} and the recent measurements performed with laser-tracked satellites.\cite{lucchesi20201} A comprehensive analysis of the Lense-Thirring effect in the solar system can be found in Ref. \citenum{Iorio2011}. The mission Gravity Probe B\cite{everitt2011gravity} was launched to measure the precession of orbiting gyroscopes.\cite{schiff1960possible} There have been other proposals, such as  LAGRANGE, which exploit 
spacecrafts located in the Lagrangian points of the Sun-Earth system,\cite{Tartaglia:2017fri} or the use of satellites around the Earth.\cite{ruggiero2019test}
In addition, we mention the GINGER experiment, which aims to measure gravitomagnetic effects in a terrestrial laboratory  by using an array of ring lasers. \cite{bosi2011measuring,ruggiero2015sagnac,di2014ring,Tartaglia:2016jfo}

Recently, the gravitomagnetic effects connected with the passage of a gravitational wave were analyized:\cite{biniortolan2017} this should not be surprising, since a gravitational wave transports angular momentum. In particular,  these effects can be easily understood by using Fermi coordinates on the basis of a gravitoelectromagnetic analogy.\cite{Ruggiero_2020} Here, we review this approach and suggest how it could be possible to detect the effects due to the magnetic-like part of a plane gravitational wave. The plan of the paper is as follows: in Section \ref{sec:fermi} we review Fermi coordinates and the definition of local spacetime metric, then we use this approach to study the effect of a plane gravitational wave in Section \ref{sec:GMR}. Conclusions are in Section \ref{sec:conc}.

\section{Local spacetime metric in Fermi Coordinates}\label{sec:fermi}

If we consider the world-line of a given observer, which ideally constitutes our laboratory frame, it is possible to write the expression of the local spacetime metric in its vicinity, using Fermi coordinates. This expression depends  both on the background spacetime and on the properties of the world-line. Fermi coordinates in the vicinity of  an arbitrary accelerated world-line  with rotating tetrads  were studied in Refs.  \citenum{Ni:1978di,Li:1979bz,marzlin}, and the general expression of the line element, up to quadratic displacements $|X^{i}|$ from the reference world-line, turns out to be
\begin{align}
ds^{2} & = -\left[\left(1+\frac{\mb a \cdot \mb X}{c^{2}} \right)^{2}-\frac{1}{c^{2}}\left(\mb{\Omega} \wedge \mb X \right)^{2}+R_{0i0j}X^iX^j \right]c^{2}dT^{2}+ \nonumber \\ & + \left[\frac{1}{c}\left(\mb{\Omega} \wedge \mb X \right)_{i}-\frac 4 3 R_{0jik}X^jX^k \right] cdT dX^{i}+ 
 \left(\delta_{ij}-\frac{1}{3}R_{ikjl}X^kX^l \right)dX^{i}dX^{j}. \label{eq:mmmetricacc}
\end{align}
Here, $\mb X$ is the position vector in the Fermi frame.  We see that in the line element (\ref{eq:mmmetricacc}) there are both the gravitational effects, deriving from the curvature tensor, and the inertial effects, due to world-line acceleration $\mb a$ and the  tetrad rotation $\bm \Omega$.  

The metric  (\ref{eq:mmmetricacc}) can be written in terms of the gravitoelectromagntic  potentials $(\Phi, \mb A)$  (see Refs. \citenum{Mashhoon:2003ax,Ruggiero_2020}), neglecting the terms $g_{ij}$ related to the spatial curvature:
\begin{equation} 
ds^2=-\left(1-2\frac{\Phi}{c^2}\right)c^{2}dT^2-\frac{4}{c}({\mb 
A}\cdot d{\mb
X})dt+\delta_{ij}dX^idX^j, \label{eq:mmetric2}
\end{equation}
where
\beq
\Phi(T, {\mb X})=\Phi^{I}(\mb X)+\Phi^{C}(T, {\mb X}),  \quad \mb A(T, {\mb X})=\mb A^{I}({\mb X})+\mb A^{C}(T, {\mb X}), \label{eq:defphiAIG}
\eeq
In particular, in the gravitoelectric potential $\Phi(T, {\mb X})$
\beq
\Phi^{I}(\mb X)=-\mb a \cdot \mb X-\frac 1 2 \frac{\left(\mb a \cdot \mb X \right)^{2}}{c^{2}}+\frac 1 2 \left[|\mb \Omega|^{2}|\mb X|^{2}-\left(\mb \Omega \cdot \mb X \right)^{2} \right] \label{eq:defphiI}
\eeq
is the \textit{inertial} contribution, while
\beq
\Phi^{C} (T, {\mb X})=-\frac{1}{2}R_{0i0j}(T )X^iX^j \label{eq:defPhiG}
\eeq
is the \textit{curvature} contribution. As for the gravitomagnetic potential  $\mb A(T, {\mb X})$, we  may distinguish the \textit{inertial} contribution
\beq
 A_{i}^{I}(\mb X)=-\left(\frac{\mb \Omega c}{2} \wedge \mb X\right)_{i}, \label{eq:defAI}
\eeq
and the \textit{curvature} contribution:
\beq
A^{C}_{i}(T ,{\mb X})=\frac{1}{3}R_{0jik}(T )X^jX^k. \label{eq:defAG}
\eeq
The gravitoelectric and gravitomagnetic fields $\mb E$ and  $\mb B$ are defined in terms of the potentials by
\begin{equation} {\mb E}=-\nabla \Phi 
-\frac{1}{c}\frac{\partial}{\partial T}\left( \frac{1}{2}{\mb
A}\right),
\quad {\mb B}=\nabla \times {\mb A}. \label{eq:defEB1}
\end{equation}
which, up to up to linear order in $|X^{i}|$, can be written as
\beq
\mb E^{I}=\mb a \left(1+\frac{\mb a \cdot \mb X}{c^{2}} \right)+\mb \Omega \wedge \left(\mb \Omega \wedge \mb X \right), \quad  E^{C}_i(T ,{\mb X})=c^{2}R_{0i0j}(T) X^j. \label{eq:defEIEG}
\eeq
and
\beq
\mb B^{I} = -\mb \Omega c, \quad B^{C}_i(T ,{\mb R})=-\frac{c^{2}}{2}\epsilon_{ijk}R^{jk}_{\;\;\;\; 0l}(T )X^l. \label{eq:defBIBG}
\eeq
In summary, the gravitoelectricmagnetic fields are written in the form
\beq
\mb E=\mb E^I+\mb E^C , \quad \mb B=\mb B^I+\mb B^C. \label{eq:defEBIG}
\eeq 
In addition, the analogy with electromagnetism can be exploited to describe the motion of free test masses; in particular, the motion of free test masses \textit{relative} to a reference mass, at rest at origin of the Fermi frame,  is determined by the geodesics of the metric (\ref{eq:mmmetricacc}). The latter can be written in the form of a Lorentz-like force equation\cite{Mashhoon:2003ax}
\beq
m\dTT{\mb X}=-m{\mb
E}-2m\frac{\mb V}{c}\times {\mb B} \label{eq:lorentz}
\eeq
up to linear order in the particle velocity $ \mb V=\dT{\mb X}$ (which is the \textit{relative velocity} with respect to the reference mass). 
Moreover, the evolution equation of classical spinning particle with spin $\mb S$ in an external 
gravitomagnetic field $\mb B$ is
\beq
\dT{\mb S}=  \frac{1}{c} \mb B \times \mb S, \label{eq:spinevol0}
\eeq 
in analogy with the corresponding equation for a charged spinning test particle in a magnetic field.\cite{Mashhoon:2003ax}

In the following Section we will apply this formalism to   plane gravitational waves.

\section{Gravitomagnetic resonance due to the passage of a gravitational wave}\label{sec:GMR}

Gravitomagnetic effects deriving from the passage of the gravitational wave can be in principle detected by using devices such as the heterodyne antenna (see e.g. Ref. \citenum{Ruggiero_2020}) or studying the perturbations of planetary motion.\cite{Iorio:2021cxt} Here, we focus on a different approach, that is based on the fulfilment of a \textit{resonance condition}.\cite{Ruggiero_2020b}

We consider a spinning particle interacting with the gravitatomagnetic field a plane wave. In the Fermi frame, where the spacetime metric is written in the form (\ref{eq:mmmetricacc}), we consider  coordinates $T,X,Y,Z$  with a set of unit vectors $\{ \mb u_{X}, \mb u_{Y}, \mb u_{Z} \}$; the direction of propagation of the wave is the  $X$ axis. In this case, for a circularly polarized wave,  the components of the gravitomagnetic field deriving from the spacetime curvature (and, hence, connected to the passage of the wave) can be written in the form\cite{Ruggiero_2020}

\begin{align}
B^{C}_{X}  &= 0, \quad B^{C}_{Y}  = -\frac{A\omega^{2}}{2}\left[- \cT Y+ \sT Z \right], \nonumber \\ 
 B^{C}_{Z}  &= -\frac{A\omega^{2}}{2}\left[\sT Y+\cT Z \right], \label{eq:defBxyz12}
\end{align}
where $A$ is the amplitude and  $\omega$  the frequency of the wave. In order to study the interaction with a spinning particle, we consider a frame  clockwise rotating  in the $YZ$ plane with the wave frequency $\omega$; then, the corresponding
basis vectors are $\mb u_{X'}=\mb u_{X}$, $\mb u_{Y'} (T)= \cT \mb u_{Y}-\sT \mb u_{Z}$ and $\mb u_{Z'} (T)= \sT \mb u_{Y}+ \cT \mb u_{Z}$. As a consequence,  the gravitomagnetic field is written as
\beq
 \mb B^{C}(T)=\frac{A\omega^{2}}{2}\left[Y\mb u_{Y'}(T)-Z\mb u_{Z'}(T)\right]
\eeq
Notice that  $ \mb B^{C}$    is a \textit{static field} in the  rotating frame that we have considered.

Let us consider the spin evolution equation  (\ref{eq:spinevol0}); the total gravitomagnetic field is $\mb B=\mb B^{I}+\mb B^{C}$, where $\mb B^{I} = -\mb \Omega c$ and it is simply proportional to the rotation rate $\mb \Omega$ of the frame. We suppose that $\mb \Omega$ is constant and it is in the direction of propagation of the wave: then,  we may write  $\mb B^{I}=-B^{I} \mb u_{X}$ where $B^{I}=\Omega c$.  As a consequence, the spin evolution equation turns out to be
\beq
\dT{\mb S}=  \frac{1}{c} \left[\mb B^{C}(T)+ \mb B^{I} \right] \times \mb S. \label{eq:spinevol1}
\eeq

{If we consider the frame co-rotating with $\mb B^{C}(T)$, since $\bm \omega=-\omega \mb u_{X}$ is the rotation rate, the time derivatives in the two frames are related by 
\beq
\dT{\mb S}= \left( \dT{\mb S} \right)_{\mathrm {rot}} +\bm \omega \times \mb S=\left( \dT{\mb S} \right)_{\mathrm {rot}}-\omega \mb u_{X'} \times \mb S. \label{eq:timederiv1}
\eeq 
Then, from Eqs. (\ref{eq:spinevol1}) and (\ref{eq:timederiv1}) we get
\beq
\left( \dT{\mb S} \right)_{\mathrm {rot}}=\left[\Delta \omega \mb u_{X'}+\frac 1 c \mb B^{C}  \right] \times \mb S =\frac 1 c \mb B_{eff} \times \mb S, \label{eq:precBeff}
\eeq
where we set $\omega-\frac 1 c B^{I}=\omega-\Omega=\Delta \omega$ and $\frac{1}{c}B^{C}=\omega^{*}$ (see below).

Then, according to Eq. (\ref{eq:precBeff}), we may say that the spinning particle undergoes a precession determined by the static effective gravitomagnetic field $\displaystyle \mb B_{eff} = c \left[ \Delta \omega \mb u_{X'} +\frac 1 c \mb B^{C} \right]$. Accordingly, we see that when  $\Delta \omega \simeq 0$, i.e.  in \textit{resonance} condition, the spin precession is around  the direction of $\mb B^{C}$, which is in any case in the $YZ$ plane, so the precession may flip the spin completely. In summary, the \textit{gravitomagnetic resonance} is obtained when the rotation rate of the frame is equal to the frequency of the gravitational wave. It is important to remember that  all precessions are referred to a reference spinning particle,\cite{biniortolan2017} at the origin of the Fermi frame so, in any case, we are talking about a \textit{relative precession.}}

Actually, the above description, which is analogous to the classical dynamics of a magnetic moment in a magnetic field, can be translated into quantum terms for a two-level system,\cite{Ruggiero_2020b} taking into account  the Hamiltonian description of the interaction of the spin of intrinsic particles in a gravitational field.\cite{Mashhoonspin,Mashhoon:2003ax,hehl1990inertial,ryder1998relativistic}   As a consequence, we may introduce a probability transition for spinning particles in the field of a gravitational wave. Let us suppose that $|g>$ and $|e>$ are the two eigenvectors, respectively of the ground and excited states,  of the projection of the spinning particle along the $X$ axis. If we suppose that a spin is, at $t=0$, in the ground state $|g>$, the probability of transition to the excited state
 $|e>$ at time $t$ is given by Rabi's formula
\beq
P_{g \rightarrow e}(T)=\frac{(\omega^{*}) ^{2}}{(\omega^{*})^{2}+\Delta \omega^{2}} \sin^{2} \left (\sqrt{(\omega^{*})^{2}+\Delta \omega^{2}} \frac T 2 \right). \label{ee:rabi1}
\eeq
Again, at resonance, i.e. when $\Delta \omega=0$, or $\omega=\Omega$, even a weak gravitational field can reverse the direction of the spin: the probability of transition is equal to 1 \textit{independently of the strength of the gravitomagnetic field}, for 
$T=\frac{2n+1}{(\omega^{*})} \pi$.

Let us add a comment on how the resonance condition could be achieved without requiring the physical rotation of our reference frame. In fact, if we consider {charged spinning particles}, we may get an equivalent situation by using a true magnetic field, on the basis of Larmor theorem, which  states the equivalence between a system of electric charges in a magnetic field and the same system rotating with the Larmor frequency.  So a  magnetic field  can be used to produce the gravitomagnetic field $\mb B^{I}$. 

\section{Conclusions}\label{sec:conc}

We have seen that, using Fermi coordinates, it is possible to emphasize the gravitomagnetic effects connected to the passage of a gravitational wave. Current detectors such as LIGO and VIRGO can detect only the interaction of a system of masses with the electric-like component of the field, so we discussed the possibility of detecting the interaction of a suitable  probe with the magnetic-like component of the wave field.
In particular, we considered the interaction of the wave with a spinning particle, both using a classical and a quantum approach, and we showed that in analogy with what happens in electromagnetism, a gravitational magnetic resonance phenomenon may appear when the reference frame rotates along the direction of propagation of the wave and the rotation rate is equal to the wave frequency. 
Actually, since it is not possible to have physical rotations for arbitrary frequencies,  we pointed out that an equivalent  situation can be obtained by using a true magnetic field, on the basis of the Larmor theorem. As for the detection of this effect, we imagine not to detect the modification of a single spinning particle, rather to consider  a great number of identical particles. For instance, the precession induced by  the gravitational wave can modify the magnetization of a macroscopical sample which, in turn, can be detected by measuring the differences in the magnetic field produced.


\end{document}